\documentclass[sigconf,authorversion,nonacm]{acmart}

\setcopyright{none}
\renewcommand\footnotetextcopyrightpermission[1]{}
\startPage{1}

\setcopyright{none}

\usepackage{amsmath, algorithm, algpseudocode}
\usepackage{algcompatible}
\usepackage{color}
\usepackage{listings}
\usepackage{multirow}
\usepackage[utf8]{inputenc}
\usepackage[T1]{fontenc}
\usepackage{microtype}
\usepackage[skip=0pt]{caption}
\usepackage{booktabs}
\usepackage[labelformat=simple,skip=0pt]{subcaption}
\usepackage{setspace}
\usepackage{graphicx}
\usepackage{graphics}
\usepackage{xspace}
\usepackage{wrapfig}
\usepackage{xspace}
\usepackage{amsfonts}
\usepackage{courier}            
\usepackage[scaled]{helvet}
\usepackage{flushend}
\usepackage{tabulary}
\usepackage{url}
\usepackage{enumerate}
\usepackage{enumitem}
\usepackage{pifont}

\usepackage{tikz}
\usepackage{ctable}
\usepackage{collcell}
\usepackage{diagbox}
\usepackage{rotating}
\usepackage{balance}

\lstdefinestyle{CStyle}{
	language=C,
	tabsize=1,
	basicstyle=\scriptsize,
	columns=fixed,
	morekeywords={loop, procedure, foreach,bool,inparallel,spawn},
	numbers=left,
	numberstyle=\tiny,
	numberblanklines=false,
	xleftmargin=10pt,
	numbersep=5pt,
	escapeinside={(*}{*)}
}

\newcommand{\cs}{$CS$}
\newcommand{\lock}{\texttt{lock()}}
\newcommand{\qnode}{\texttt{qnode}}
\newcommand{\unlock}{\texttt{unlock()}}
\newcommand{\rlock}{\texttt{RLock()}}
\newcommand{\runlock}{\texttt{RUnlock()}}
\newcommand{\wlock}{\texttt{WLock()}}
\newcommand{\wunlock}{\texttt{WUnlock()}}
\newcommand{\rel}{\texttt{release()}}
\newcommand{\acq}{\texttt{acquire()}}

\newcommand{\uu}{unbalanced-unlock}

\newcommand{\misbehaver}{$T_m$}
\newcommand{\cmark}{\ding{51}}%
\newcommand{\xmark}{\ding{55}}%
\newcommand{\succwt}{\texttt{succ\_must\_wait}}%
\newcommand{\false}{\texttt{false}}%
\newcommand{\true}{\texttt{true}}%
\newcommand{\cas}{\texttt{CAS}}%
\newcommand{\swap}{\texttt{SWAP}}%
\newcommand{\pid}{PID}%
\newcommand{\NULL}{\texttt{NULL}}%

\newcommand*{\MinNumber}{0.0}%
\newcommand*{\MidNumber}{5} %
\newcommand*{\MaxNumber}{120}%

\newcommand{\ApplyGradient}[1]{%
        \ifdim #1 pt > \MidNumber pt
            \pgfmathsetmacro{\PercentColor}{max(min(100.0*(#1 - \MidNumber)/(\MaxNumber-\MidNumber),100.0),0.00)} %
            \hspace{-0.33em}\colorbox{red!\PercentColor!yellow}{#1}
        \else
            \pgfmathsetmacro{\PercentColor}{max(min(100.0*(\MidNumber - #1)/(\MidNumber-\MinNumber),100.0),0.00)} %
            \hspace{-0.33em}\colorbox{green!\PercentColor!yellow}{#1}
        \fi
}
\newcolumntype{R}{>{\collectcell\ApplyGradient}c<{\endcollectcell}}

\begin{document}
\title{Protecting Locks Against Unbalanced \texttt{Unlock()}}
\author{Vivek Shahare}
\affiliation{%
  \institution{Indian Institute of Technology} \city{Dharwad}
  \country{India}}
\email{vivek.shahare@iitdh.ac.in}

\author{Milind Chabbi}
\affiliation{%
  \institution{Programming Systems Group \\ 
  Uber Technologies Inc.}
  \city{Sunnyvale}
  \state{CA}
  \country{USA}
}
\email{milind@uber.com}

\author{Nikhil Hegde}
\affiliation{%
  \institution{Indian Institute of Technology} \city{Dharwad}
  \country{India}}
\email{nikhilh@iitdh.ac.in }

\begin{abstract}
The lock is a building-block synchronization primitive that enables mutually exclusive access to shared data in shared-memory parallel programs. 
Mutual exclusion is typically achieved by guarding  the code that accesses the shared data with a pair of  \texttt{lock()} and \texttt{unlock()} operations.
Concurrency bugs arise when this ordering of operations is violated.

In this paper, we  study a particular pattern of misuse where an \texttt{unlock()} is issued without first issuing a \texttt{lock()}, which can happen in code with complex control flow.
This misuse is surprisingly common in several important open-source repositories we study.
We systematically study what happens due to this misuse in several popular locking algorithms.
We study how misuse can be detected and how the locking protocols can be fixed to avoid the unwanted consequences of misuse.
Most locks require simple changes to detect and prevent this misuse. 
We evaluate the performance traits of modified implementations, which show mild performance penalties in most scalable locks.

\end{abstract}




\maketitle

\section{Introduction}
Locks are synchronization primitives widely used to ensure mutual exclusion in shared-memory parallel programs. 
Locks remain the most popular general-purpose mutual exclusion primitives; the optimistic concurrency~\cite{KungOCC} alternatives such as transactional memory~\cite{TMLecture,HerlihyTM, WangIBMBGQ, DiceTMEarly, YooIntelTSX} have had limited success~\cite{McKenneyLockvsTM, YooTMTires, larabel2021intel} in parallel computing.

The emergence of many/multicore systems~\cite{OlukotunCMP}, growing layers of memory hierarchy~\cite{jacob2009memory}, and on-chip communication~\cite{jerger2017chip} medium has spurred the creation of numerous locking protocols. 
A majority of the existing work on locks~\cite{chabbi2015high, guerraoui2019lock, dice2012lock, tocs91, lozi2016fast, Boyd-wickizerNSL,calciu2013numa, ChabbiAHMCS,DiceMalthusian, KashyapShuffle,RadovicHierarchicalBOLock, hemlock, DiceCNA} focuses on the design, performance, and suitability of the locks on various multicore architectures. 
In this paper, we focus on the misuse scenarios i.e., what happens on {\em accidental} incorrect use of locking operations.

Any locking protocol provides at least the following two operations on the \texttt{mutex} object --- \texttt{lock} and \texttt{unlock}\footnote{ 
Additional operations may include \texttt{trylock} or arguments to \texttt{lock} that control the wait time~\cite{chabbiHMCST, ScottNBAbort, ScotTimeout}.A \texttt{synchronized}  code region is mutex protected in Java.}.
We refer to the region guarded by a pair of \texttt{lock-unlock} operations as a critical section (\cs{}).
Lock misuses can be of the following broad categories:
\begin{enumerate}[leftmargin=*]
\item \textbf{Unbalanced-lock problem}: Here, a thread forgets to unlock at the end of \cs.
This is a common problem, which results in the starvation of all other threads attempting to acquire the same lock. This situation does not corrupt the system; it may hurt performance and/or halt the progress of the system. 
It is relatively easy to observe via performance analysis tools~\cite{ShendeTau, AdhiantoHPCToolkit, TallentLock, TallentMTApps, LiuOpenMP}.

A special case of this problem happens when the same thread tries to re-acquire the lock it already holds.
If the lock is reentrant~\cite{haack2008reasoning}, 
subsequent re-acquisition succeeds immediately.
If the lock is non-reentrant, it results in a deadlock.

\sloppy
\item  \textbf{Unbalanced-unlock problem}. Here, a thread tries to \unlock{} without holding the lock.
Unbalanced unlock may result in admitting more than one thread into the \cs{}, violating the mutual exclusion and hence can lead to undefined program behavior. 
It can be harder to debug.
Some locks can corrupt the lock internals. 
Different locking protocols lead to different behaviors.
Some locking algorithms may easily exhibit  an egregious program behavior (e.g., a Ticket lock), whereas some other locking algorithms may require subtle thread-interleaving to expose the problem (e.g., Graunke-Thakkar lock~\cite{GraunkeThakkarLock}) while some others may be nearly immune to \uu{} (e.g., HCLH lock~\cite{luchangco2006hierarchical}).
Hence, \textbf{this is the topic of focus of this paper.}

Reentrant and reader-writer locks represent special cases of these unbalanced-unlock issues. For example,  a thread may perform more \unlock{}s than the corresponding number of \lock{} operations on a reentrant lock.
In reader-writer locks, a read-only \lock{} may be incorrectly paired with a read-write \unlock{} or vice-versa.
\end{enumerate}

Listing~\ref{lst:unbalUnlock} shows an example of \uu{} bug in the Linux kernel. In this example, the buggy code jumps to \texttt{out}  label if the flag \texttt{wilc->quit} is \texttt{true} (line 2). This causes the \texttt{mutex\_unlock()} to be called without acquiring the lock first.

\begin{figure}[!t]
\begin{lstlisting}[language=C++, caption=Unbalanced unlock in Linux at \texttt{drivers/staging/wilc1000/wlan.c} (sha: \texttt{bd4217c})., label=lst:unbalUnlock]
...
    if (wilc->quit){
        goto out;
    }
    mutex_lock(&wilc->txq_add_to_head_cs);
    tqe = wilc_wlan_txq_get_first(wilc);
    if (!tqe){
        goto out;
    }
...
out:
    mutex_unlock(&wilc->txq_add_to_head_cs);
	    
    *txq_count = wilc->txq_entries;
    return ret;
...
\end{lstlisting}
\end{figure}

Several questions arise in the context of unbalanced-unlock for a given locking algorithm.

\begin{itemize}[leftmargin=*]
    \item can it violate mutual exclusion? if so, how many threads may enter the \cs{} simultaneously? 
    \item can it prevent some other thread from ever acquiring the lock? if so, how many threads may starve? 
    \item can a lock {\em recover} from  misuse? i.e., could there be any benign cases not causing any side effects on the lock? 
    \item can this be detected/prevented and if so, how? 
\end{itemize}

In this paper, we focus on {\em spinlocks} only and present a systematic study of popular spin-lock protocols to find answers to the above questions. 
We modify the lock internals to prevent misuse and evaluate the modified implementations to measure performance.  We make the following contributions in this paper:
\begin{itemize}[leftmargin=*]
    \item Show that \uu{} is a surprisingly common lock misuse scenario in many critical systems.
    \item Systematically study the behavior of popular locking algorithms in the event of an \uu{}.
    \item Provide remedies to popular locks to make them resilient to the \uu{} situation.
    \item Empirically demonstrate that making popular scalable locks resilient to \uu{} need not significantly compromise their performance while retaining the original design goals.
\end{itemize} 

The rest of the paper is organized as follows.
We motivate the need to study the \uu{} problem in  Section~\ref{sec:motivation}; we study the behavior of some  popular locks,   reader-writer,  and software-only locks in Section~\ref{sec:popular}-\ref{sec:sw}; we evaluate the performance of locks after modifications in Section~\ref{sec:Evaluation}; we present the related work in Section~\ref{sec:related} and offer our conclusions in Section~\ref{sec:conclusion}.

\sloppy
In the rest of this paper, we use the terms \lock{} and \acq{} synonymously and \unlock{} and \rel{} synonymously. 
We use the term ``mutex'' to mean mutual exclusion and ``misuse'' to mean unbalanced-unlock where the context is obvious.
\cas{} refers to atomic compare-and-swap and \swap{} refers to atomic exchange operation.

\section{Motivation for studying \uu{}}
\label{sec:motivation}

\subsection{Prevalence of \uu{}}
 We investigated several large open-source repositories to quantify the gravity of the unbalanced-unlock problem, especially in relation to the unbalanced-lock problem.
 For our study, we chose popular, large, open-source code repositories covering different domains: 
 concurrent programming language, runtime, and  standard libraries (Golang);
  operating system (The Linux);
  developer tools (LLVM);
  database systems (MySQL);
  concurrent high-performance cache (memcached).
  We examined the entire code commit history looked up for strings such as {\small unlock, mutex, double unlock, unlock without lock, lock placement, deadlock, starvation, improper, release lock, lock misuse, missing lock, missing unlock, stray unlock, forget to unlock, holding lock, without acquiring, without unlocking, acquiring the lock, and forgetting to release a lock}. From the search results, we excluded the ones that indicated code changes pertaining to lock design and performance; we studied those pertaining to bugs and potential bugs of the aforementioned categories. 
  
\begin{figure}[t!]
	\centering
	\includegraphics[scale=0.5]{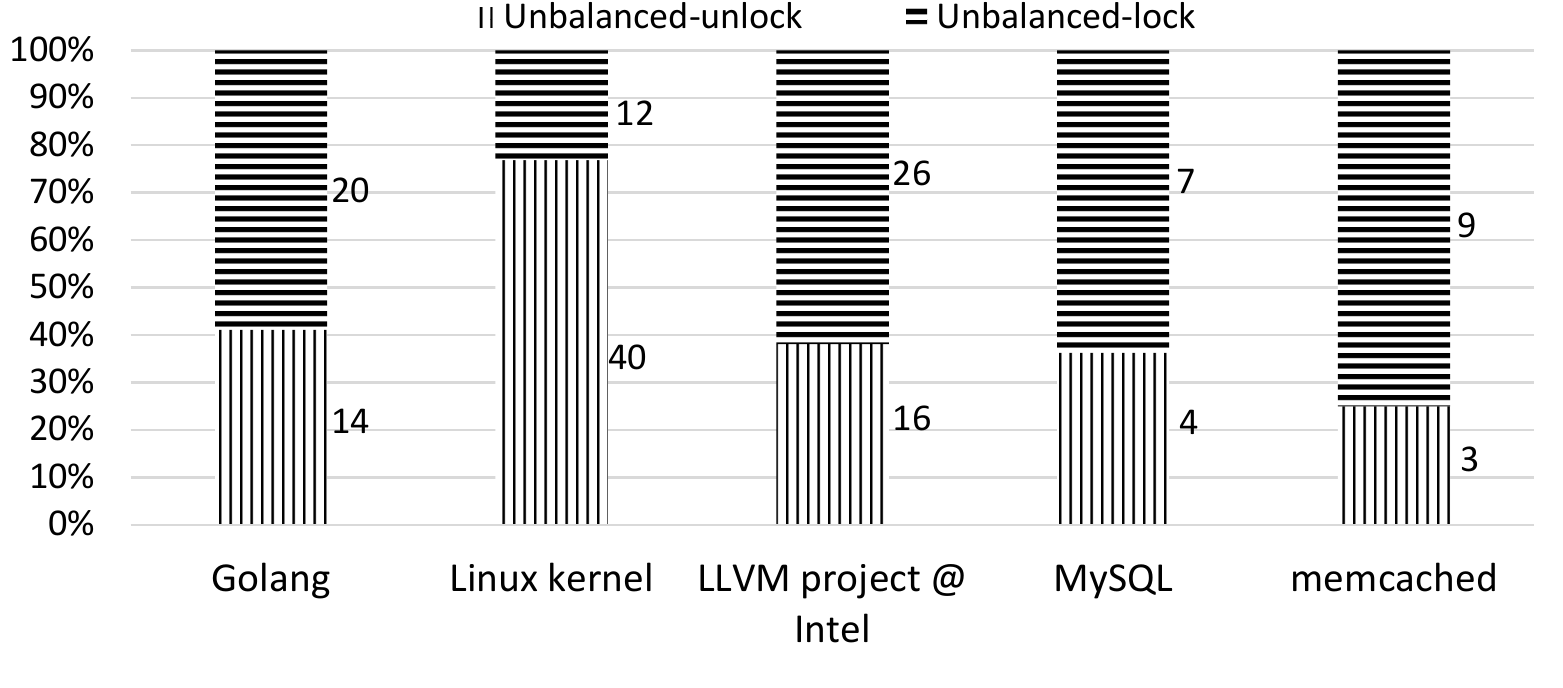}
	\caption{Lock related code changes (commits) in large open-source projects categorized by misuse type.}
	\label{fig:motivation}
\end{figure}

  Figure~\ref{fig:motivation} shows that the unbalanced-unlock is surprisingly common and significant.
In this figure \emph{unbalanced-lock} category covers: forgetting to release a lock, inability to release a lock for some reason e.g., destroyed mutex, acquiring the same lock while the lock is already acquired by the same thread, incorrect placement of the lock acquire and release methods. The \emph{unbalanced-unlock} category covers: incorrectly releasing a lock when the lock is not acquired, including releasing the lock twice and unbalanced reader-writer locks. The numbers indicate the misuse instances corresponding to the two scenarios. 
While the numbers vary from application to application.\footnote{Of the linux kernel bugs shown in Figure ~\ref{fig:motivation}, 62.5\% were found in drivers (such as IEEE802.11 and ethernet), 10\% each in Filesys and Memory modules, 5\% each in Network, Process, and Platform, and 2.5\% in IO/storage module. Some of these bugs were recent (2022), and some were old (2012 since the bug database is maintained).} 
, it is evident that \uu{} is both common and a significant fraction of this category of bugs.

\subsection{The role of the locking algorithms}
What happens to the system/application in the event of an \uu{} depends on numerous considerations such as: 
\begin{enumerate}[leftmargin=*]
    \item whether or not the lock results in violating the mutual exclusion?
    \item  if a data race ensues due to mutex violation, whether the programming model/language offers any semantic guarantees or does it lead to the so-called ``catch-fire'' semantics (e.g., C++~\cite{boehmcppmodel})?
    \item whether or not one or more threads get into starvation?
    \item whether or not the lock internals be corrupted in a way that subsequent lock usage is hampered?
\end{enumerate}

Answering some of these questions are application context and programming model related and hence outside the realm of the locking algorithm. However, the locking algorithm has a crucial role to play in mutex violation or starvation.

What a locking algorithm does in an \uu{} is not a simple answer because lock algorithms come in myriad shapes and forms.
Some allow FIFO property, some do not; 
some enable read-only feature, some do not;
some use hardware-provided atomics and others are software-only locks; some are tailored for the memory hierarchy while others are not; some allow recursive acquisitions others do not; some need to carry a context from acquire to release while other do not; to a few varieties.

Some locks such as the Array-based queuing locks~\cite{anderson1990performance} cause mutex violation but do not cause starvation; locks such as Graunke-Thakkar lock~\cite{GraunkeThakkarLock} cause only starvation but not mutex violation; locks such as the MCS lock~\cite{tocs91} can cause both and additionally may cause illegal memory access.
A clear understanding of the lock's behavior in the event of a misuse helps us a) assess the potential worst-case scenario and b) modify the lock internals such that it can self-detect that an invocation of \rel{} is an instance of \uu{} and avert potential problems.  

\subsection{Generic solution and its insufficiency}
Any lock implementation can be enhanced in a generic manner to detect \uu{} by maintaining an additional \pid{} field in the lock.
Such \pid{} would be set to the lock acquiring thread's id on lock acquisition; the \rel{} would check the stored \pid{} against its own \pid{}; in case of a mismatch flag an error and return.

Naively introducing  a \pid{} field, however, may not be advisable under all circumstances for the following reasons:
\begin{enumerate}[leftmargin=*]
    \item The \pid{} field increases the memory footprint of the lock, which has significant consequences when millions of lock instances are created and used concurrently\footnote{The $fluidanimite$ application in PARSEC3.0~\cite{bienia11benchmarking} benchmark has 1.7 million lock instances.}.
    \item Several systems, such as the Linux Kernel, require the lock to be compact, sometimes only a single machine word~\cite{WangMCSGuest,DiceCNA}.
    \item A \pid{} may be unavailable in some task-based systems (e.g., Golang~\cite{nogoid} and CilkPlus~\cite{cilkPluswww}) since tasks may migrate across OS threads.
\end{enumerate}

Our attempt here is to avoid using a \pid{} as much as possible, and we fall back to the \pid{} solution only when unavoidable. 

The details of misuse scenarios, their side effects, and handling the side effects are the subject of the next three Sections.  

\section{Unbalanced Unlock in Popular Locking Algorithms}
\label{sec:popular}
In this section, we study popular locking algorithms and what happens on an instance of \uu{}.
We also propose detectability and remediation as code deltas to the pseudo-code of the original locks.
 We refer the reader to~\cite{tocs91, dice2012lock, chabbi2015high, magnusson1994queue, craig1993building} for the basic knowledge of test-and-set, ticket, array-based queueing, MCS, CLH, and hierarchical locks.
We refer to the misbehaving thread as \misbehaver{}.
We adapt the pseudo-code from ~\cite{tocs91} by Mellor-Crummey and Scott for our discussions but use the C++11 syntax in our discussion.
We use the C++11 \texttt{atomics} for all shared variables and assume sequential consistency as the memory model.
We modify the signature of all \rel{} functions to return \false{} in the event of unbalanced-unlock and \true{} otherwise. 

Our work is centered around making minimal modifications to the locking protocols to detect and remedy \uu{}.
There are other ways in which the lock can be misused---for example, passing a null or arbitrary lock pointer to \acq{} or \rel{} APIs.
Furthermore, locks such as MCS, CLH, and Array-based queuing locks, need thread-local ``context'' pointer that is carried from the \acq{} API to the \rel{} API.

Any time a pointer is passed to an API, there is the potential to cause illegal memory access.
We cannot possibly fix this issue in an unsafe language such as C. 
For languages such as C, our algorithmic modifications only detect and fix the misuse if the context pointer is a legal memory owned by the thread.
An arbitrary pointer is not possible in memory-safe languages such as Java---a null-pointer check suffices.
Our solution to avoid arbitrary pointers in C++ is to pass objects by  lvalue references~\cite{cpprefAlias} instead of pointers; C++ references can only point to valid objects unless someone takes the additional burden to circumvent this.
All our code examples follow this strategy.
In what follows, we, typically, do not discuss the consequences of passing a null or a rogue lock/context pointer to the \rel{} API as part of \uu{}.  

\subsection{Test and Set (TAS) lock}
\label{sec:tas}
\begin{figure}
	\centering
	\includegraphics[scale=0.80]{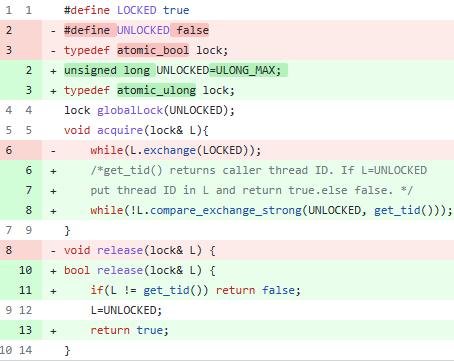}
	\caption{Test and Set lock before and after the fix.}
	\label{fig:tas}
\end{figure}

The TAS lock (Figure~\ref{fig:tas}) is the simplest lock.
There is a shared boolean variable \texttt{L} initialized to \texttt{UNLOCKED}.
The \acq{} protocol attempts to \swap{} \texttt{L} to \texttt{LOCKED}  value and enters the \cs{} on success or retries on failure. 
The \rel{} protocol unconditionally resets \texttt{L} to \texttt{UNLOCKED}.
The details discussed herein apply to the family of TAS locks, which include Test-and-test-and-set (TATAS) and TATAS with exponential back-off (BO)~\cite{tocs91}.

Different situations arise in the \uu{} of a TAS lock.
If the lock is not held by another thread, when the misbehaving thread, \misbehaver{}, resets \texttt{L} in \uu{}, there is no visible effect on the entire system.
If the lock is, however,  already held by another thread and there are waiting threads, resetting \texttt{L} to \texttt{UNLOCKED} by \misbehaver{} can admit an additional waiting thread into the \cs{} violating the mutual exclusion.
Every instance of an \uu{} can admit at most one additional thread simultaneously into the \cs{}. Put alternatively, $N$ instances of unbalanced unlocks can admit at most $N$ threads simultaneously into the \cs{}.
The misuse does not introduce any starvation---neither the misbehaving thread nor other threads spin forever; note, however, that, by design, the TAS lock does not ensure starvation freedom to any waiting thread.

\textbf{Detection and solution:}
We can detect an attempt to perform an \uu{} by storing the \pid{} into \texttt{L} instead of a boolean.
Figure~\ref{fig:tas} shows the algorithm for acquiring and releasing the TAS lock and the suggested changes to prevent the side effects of unbalanced unlock operation.

In the modified form, the \acq{} protocol employs a \cas{} to set the thread's \pid{} at \texttt{L}.
During \rel{}, it checks if the releasing thread's \pid{} matches the value stored at \texttt{L}. A mismatch indicates \uu{} and we skip resetting the value pointed to by \texttt{L} to \texttt{UNLOCKED}.
Our solution does not introduce a new field in the lock to store the \pid{}; it re-purposes the same lock word with the \pid{}.  

\subsection{Ticket lock}
\label{sec:tkt}
\begin{figure}
	\centering
	\includegraphics[scale=0.85]{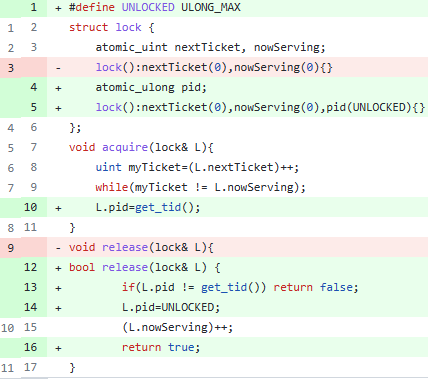}
	\caption{Ticket lock before and after the fix.}
	\label{fig:ticketlock}
\end{figure}

\sloppy
The ticket lock (Figure~\ref{fig:ticketlock}) is a FIFO lock.
It employs two global variables---\texttt{nowServing} and \texttt{nextTicket}. Each thread wanting to enter the \cs{} atomically increments \texttt{nextTicket} to gets its position in the queue and waits until \texttt{nowServing} equals its position. The exiting thread increments \texttt{nowServing}. In the event of an \uu{}, the following problems arise.

\sloppy
\textbf{Mutex violation:}
If the misbehaving thread \misbehaver{} increments \texttt{nowServing} when one thread ($t_1$) is already in the \cs{}, it admits the successor thread ($t_2$) into the \cs{} violating the mutual exclusion.
The situation beyond this is quite complex. 
Without the loss of generality, assume that $t_1$ finishes its \rel{} before $t_2$ calls its \rel{}.
This admits the next waiting thread ($t_3$) into the \cs{}, along with the already present $t_2$ and the situation of two threads in the \cs{} can propagate.
Alternatively, $t_1$ and $t_2$ may racily update \texttt{nowServing} from a value of, say $n$, to only $n+1$, making the lock recover for the remaining threads.
A single instance of \uu{} will allow at most two threads to be simultaneously into the \cs{}.
$N$ instances of \uu{}s will allow at most $N+1$ threads to be simultaneously into the \cs{}.

\sloppy
\textbf{Starvation}:
Starvation ensues if \texttt{nowServing} does not increase monotonically.
For example: let \texttt{nextTicket=nowServing=x}.
Now let, \misbehaver{} load \texttt{nowServing}; 
let an arbitrary number $n>1$ of \texttt{lock-unlock} pairs finish making \texttt{nextTicket=nowServing=x+n}.
Now let, \misbehaver{} store \texttt{nowServing=x+1}.
After this, all new lock requests will starve since 
\texttt{nowServing} will never reach \texttt{myTicket} value for any thread.

In almost all cases, a misbehaving thread in a ticket lock causes starvation for all other threads, violation of mutual exclusion, or both.  The only time when there is no detrimental effect is when the lock holder and \misbehaver{} both racily update \texttt{nowServing} to the same value and no other thread notices the updated \texttt{nowServing} in the intervening window. 

Interestingly, the misbehaving thread itself does not experience starvation, unless it attempts to \acq{} later.

\textbf{Detection and solution:}
By introducing a \texttt{pid} field into the lock as shown in the Figure~\ref{fig:ticketlock}, we remedy the lock.
The \texttt{pid} is set after lock acquisition; the \rel{} checks to make sure the releasing thread id matches the \texttt{pid}, otherwise, it does not update \texttt{nowServing}.

\subsection{Array-based Queuing Locks (ABQL)} 
\subsubsection{Anderson's lock:}
\label{sec:abql}

\begin{figure}
	\centering
	\includegraphics[scale=0.85]{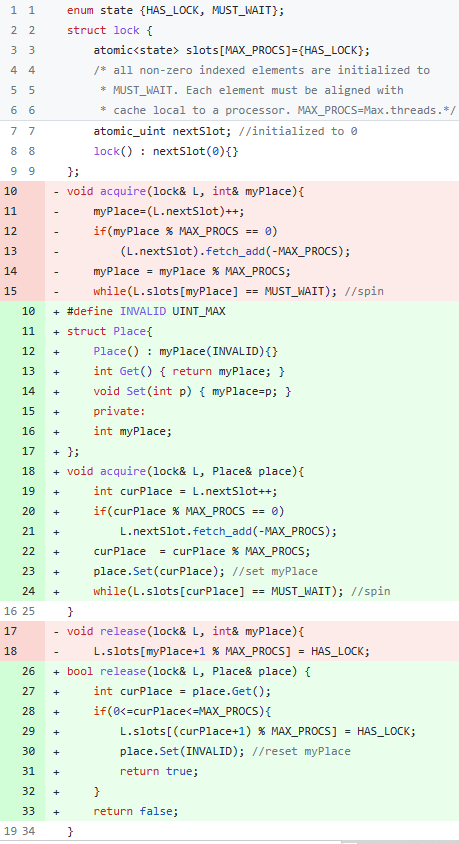}
	\caption{Anderson's lock before and after the fix.}
	\label{fig:arraylock}
\end{figure}

Anderson's array-based lock~\cite{anderson1990performance} (Figure~\ref{fig:arraylock}) is also a FIFO lock.
In a nutshell, there is an array \texttt{slots} of \texttt{MAX\_PROCS} size.
A thread desiring to acquire the lock obtains a slot (indexed by \texttt{myPlace}) in this array via an atomic add (\texttt{fetchAndAdd}) operation and spins on its slot.
The lock-releasing thread unblocks the next waiting thread in the array.

\texttt{myPlace} is a crucial variable here.
It is the per-thread context that is carried from \acq{} to \rel{}.
An \uu{} can pass an uninitialized value of \texttt{myPlace} to \rel{}.
The modulus operation in \rel{} protects against array out-of-bounds accesses.
However, it can release a waiting thread into the \cs{} while there might already be another thread in the \cs{}. The effect can cascade as the two threads now in the \cs{} can release two more threads into \cs{} during their \rel{}. 
There will not be any starvation of any thread---the modulus operation acts as a safety guard.

\textbf{Detection and solution: } We wrap the \texttt{myPlace} integer into an object (\texttt{Place}) that is always initialized to a special value (\texttt{INVALID}) via the object constructor and make it inaccessible to the user code via C++ private field, as shown in the Figure~\ref{fig:arraylock}.
\texttt{myPlace} is set to a legal value after the lock is acquired and reset to \texttt{INVALID} after the lock release. 

\subsubsection{Graunke-Thakkar lock (GT-Lock):}
\label{sec:gtlock}

\begin{figure}
	\centering
	\includegraphics[scale=0.8]{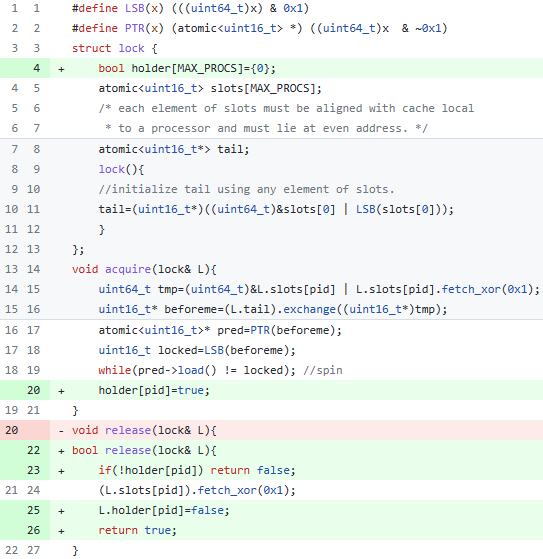}
	\caption{Graunke \& Thakkar's lock.}
	\label{fig:arraylock_gnt}
\end{figure}

GT-Lock~\cite{GraunkeThakkarLock} (Figure~\ref{fig:arraylock_gnt}) is also a FIFO lock and organizes the threads in a queue. However, unlike ABQL, the waiting threads are on disjoint memory locations (similar to the CLH lock~\cite{magnusson1994queue, craig1993building}). 
The global \texttt{lock} structure has two fields---an array of booleans \texttt{slots} and a \texttt{tail} pointer.
The \texttt{slots} is organized such that each element falls in a processor-private memory module and is on an even-numbered address---the lowest-order bit (LSB) zero (the code uses \texttt{uint16\_t} for this purpose); the elements are uninitialized; and each thread owns a unique element identified by its \pid{}.

The \acq{} for a thread with id \pid{} publishes its slot address and its value (encoded via bitwise ORing \texttt{\&slots[pid]|slots[pid]}) at the \texttt{tail} via the \swap{} operation.
In return, it obtains its predecessor's location ( \texttt{pred}) and the value (\texttt{locked}) to wait on.
The thread then spins until the value at address \texttt{pred} is \texttt{locked}. Correspondingly, the \rel{} protocol toggles the value at \texttt{slots[pid]} to pass the lock to a waiting successor via an atomic \texttt{xor}.

To bootstrap the process, \texttt{tail} is initialized to point to the address of an arbitrary element of the slot (say \texttt{\&slots[0]} element) and bitwise ORed with the logical negation of the value at such location ($\neg$\texttt{slots[0]}).

In the case of an \uu{}, the GT-lock does not violate the mutual exclusion but can cause starvation for all other threads. 

\textbf{No mutex violation:} Mutex violation is not possible because of the following argument. The slot \texttt{slots[}\misbehaver\texttt{]} that the \misbehaver{} toggles, is either in the queue or not in the queue; if it is not in the queue, the bit toggling has no impact on any thread.
If it is in the queue, it is either at the head of the queue or not at the head of the queue.
If it is at the head of the queue, then no thread is in the \cs{}, toggling in the \rel{} due to \uu{} will only release a legit successor (say $T_s$) into the \cs{}.
Finally, \texttt{slots[}\misbehaver\texttt{]} may not be at the head of the queue if-and-only-if \misbehaver{} enqueued it, which can happen only via an \acq{} operation, which contradicts the fact that the \rel{} is an \uu{}.

\textbf{Starvation:} The \misbehaver{} can cause starvation in the following way: consider \misbehaver{} performing a round of successful \texttt{lock-unlock} and during this time a successor $T_s$ starts spinning on \texttt{slots[}\misbehaver{}\texttt{]}. 
Before $T_s$ observes the change of the bit,  \misbehaver{} may perform another \rel{}, causing an instance of \uu{} reverting the bit back to the original value. Thus $T_s$ will miss the update and may wait indefinitely. Due to the FIFO property, no thread will enter the \cs{}. 

\textbf{Detection and solution:}
We introduce another boolean array \texttt{holder} of size \texttt{MAX\_PROCS} and initialize all elements to \false{}.
We change the \acq{} protocol to set \texttt{holder[pid]} to \true{} after lock acquisition.
The releaser checks if \texttt{holder[pid]} is \true{} and detects \uu{} if is \false{}. 
A successful \rel{} protocol resets this bit to \false{}. 
We note that one can repurpose the \texttt{slots} array without having to create an additional \texttt{holder} array.

\subsection{MCS lock} 
\label{sec:mcs}

\begin{figure}
	\centering
	\includegraphics[scale=0.70]{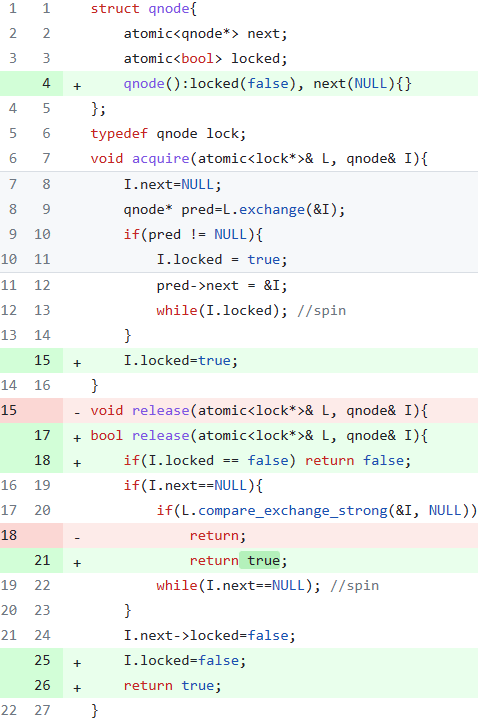}
	\caption{MCS lock before and after the fix.}
	\label{fig:mcs}
\end{figure}

The MCS lock (Figure~\ref{fig:mcs}) is a FIFO lock.
It forms a singly linked list of waiting threads, each spinning on its own flag (\texttt{locked} field) in a \qnode{} structure. Each thread enqueues itself into the queue by \swap{}ing the address of its \qnode{} with the tail pointer \texttt{L}. This operation allows it to know its predecessor. An additional step of updating the predecessor's \texttt{next} pointer is involved before starting to spin locally.
The release protocol involves one of the following: a) following the \texttt{I.next} and setting the successor's \texttt{locked} to \false{}, or b) \cas{}ing tail pointer to \NULL{} if there is no successor.

For the MCS lock, in the event of an \uu{}, the following situations can happen:
\begin{enumerate}
    \item if \texttt{I.next} is null, the misbehaving thread will loop forever  waiting for a successor to enqueue.
    \item if \texttt{I.next} is an arbitrary pointer, it can corrupt the memory leading to unpredictable behavior.    
    \item if \texttt{I.next} points to another legal \qnode{} that is already enqueued, it will release such thread into \cs{} violating the mutual exclusion. This situation arises when passing a previously used \qnode{} to the \rel{} operation during \uu{}. In this case, a previously set \texttt{I.next} may be pointing to a \qnode{} that is also enqueued again.
\end{enumerate}

Note that the C++ lvalue reference avoids \texttt{I} from taking a rogue or a null value.

The MCS lock does not cause starvation for threads other than \misbehaver{}. The argument for no starvation for other threads is the following: there are two loops one in \acq{} and another in \rel{}.
A misbehaving thread can only set the \texttt{locked} field to \false{}, releasing any waiting thread in \acq{}.
The loop in \rel{} is on the \texttt{next} field, which is untouched by a misbehaving thread.

\textbf{Detection and solution:}
Our solution to the MCS lock is to always set the \texttt{locked} to \true{} after the lock acquisition. In the \rel{}, we reset \texttt{locked} to \false{}. With this scheme, we can detect a misplaced \rel{} because \texttt{I.locked} will be \false{}.

\begin{figure}
	\centering
	\includegraphics[scale=0.85]{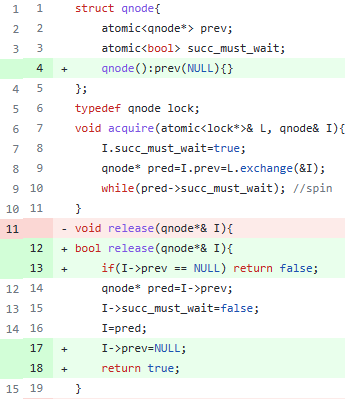}
	\caption{CLH lock before and after the fix.}
	\label{fig:clh}
\end{figure}
\subsection{CLH lock} 
\label{sec:clh}

The CLH lock~\cite{magnusson1994queue, craig1993building} (Figure~\ref{fig:clh}) is like the MCS lock with subtle differences.
It is also formed of the linked list of \qnode{}s.
However, instead of each \qnode{} maintaining the next pointer, it maintains a pointer to the predecessor \texttt{prev}.
Each thread brings its own \qnode{} \texttt{I} and enqueues it into the linked list pointed to by the tail pointer \texttt{L}. However, the spin wait is not a flag in its node (\texttt{I}), but instead on the flag (\succwt{}) of the predecessor.
In the release protocol, the lock owner toggles the flag {\texttt{I.succ\_must\_wait}} to pass the lock to the successor; the releaser then takes the ownership of its predecessor \qnode{}, which it may reuse in another locking episode or free it.
To bootstrap the process, the  tail pointer \texttt{L} of the CLH lock initially points to a dummy \qnode{} whose  \succwt{} flag is already set to \false{}.

\begin{figure}[!t]
    \centering
    \begin{subfigure} {\linewidth}
        \centering
        \includegraphics[width=\linewidth]{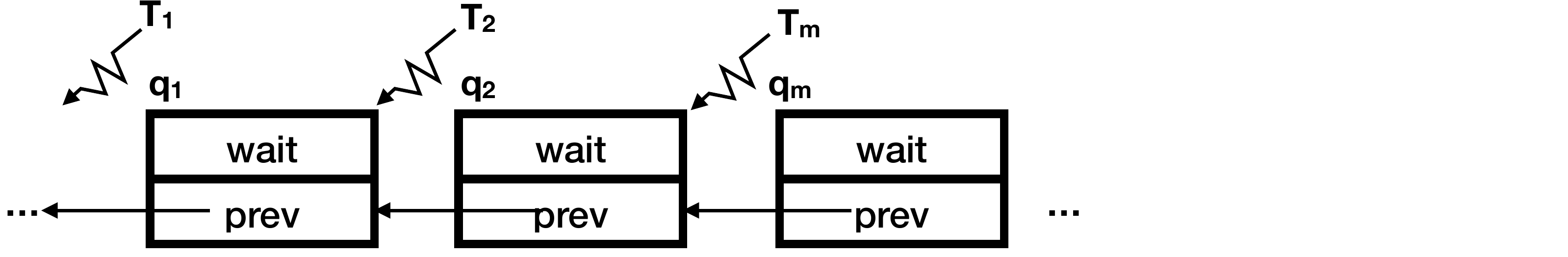}
        \caption{One episode of successful locking.}
        \label{fig:clhinit}
    \end{subfigure}
    \begin{subfigure} {\linewidth}
        \centering
        \includegraphics[width=\linewidth]{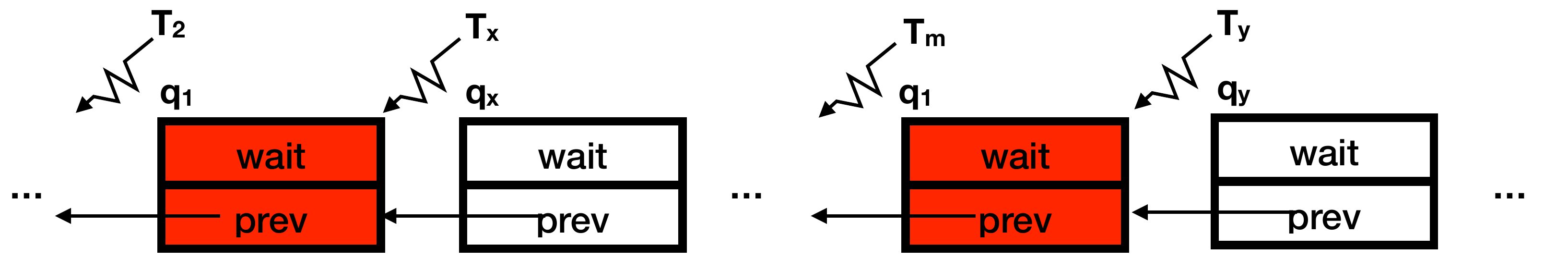}
        \caption{Re-enqueue after \uu{}.}
        \label{fig:clhviolate}
    \end{subfigure}
    \caption{Starvation and Violation of Mutex in CLH.}
    \label{fig:chlmisuse}
\end{figure}

\textbf{Mutex violation:} in the event of an \uu{} in a CLH lock, if \texttt{I} points to a legit \qnode{} waited to on by another thread (say $T_2$) while some thread (say $T_1$) is already in the CS, it can result in letting $T_2$ in the CS, thus violating the mutual exclusion.

The second situation is quite likely to happen in the CLH lock because the releasing thread gets the ownership of the predecessor node. 
Consider the initial setting where three threads $T_1$, $T2$, and \misbehaver{} are enqueued in sequence as shown in Figure~\ref{fig:clhinit}. They respectively enqueue \qnode{}s $q_1$, $q_2$, and $q_m$. In this episode of locking, they are all well-behaved. 
At the end of this round, $T_2$ takes ownership of $q_1$ and \misbehaver{} takes the ownership of $q_2$.
Now, let, \misbehaver{} perform an \uu{}.
As a result, it incorrectly assumes the ownership of $q_1$---the predecessor of $q_2$.

Now, let, both $T_2$ and \misbehaver{} participate in another round of locking.
Let $T_2$ enqueue $q_1$, followed by another thread, say $T_x$. Subsequently, let \misbehaver{} also enqueue $q_1$, followed by another thread $T_y$ as shown in Figure~\ref{fig:clhviolate}.
Let, the lock be passed to $T_2$.
During the lock release, $T_2$ will set $q_1$'s \texttt{succ\_must\_wait} to \false{}, resulting in both $T_x$ and $T_y$ to enter the CS, thus violating the mutual exclusion.

\textbf{Starvation:} Continuing with the previous example: since both $T_2$ and $T_m$ access the fields of $q_1$ in Figure~\ref{fig:clhviolate} it will cause a data race.
The racy updates, can result in corrupting the linked list (making it cyclic) and moreover, it can cause the update to the \texttt{succ\_must\_wait} field to be lost. As a result, the lock may never be passed to any successor resulting in starvation.

\textbf{Detection and solution:}
It is evident from the previous problems, that the ability of the \uu{} to get access to an arbitrary \qnode{}, via the \texttt{prev} pointer is the source of all issues. The remedy is to prevent  \texttt{prev} from continuing to point to any \qnode{} once the \rel{} is done.

In our solution (Figure~\ref{fig:clh}), at the end of the \rel{}, we reset the \texttt{prev} pointer of \texttt{I} to \NULL{}. This modification is outside the critical path.
Additionally, in our C++ constructors, we initialize the \texttt{prev} field of a \qnode{} to \NULL{} via the \qnode{} object constructor and check in the \texttt{release} method if the \texttt{prev} value is \NULL{}.

\subsection{MCS-K42}
The MCS-K42~\cite{mcsk42} lock eliminates  allocating and passing of \texttt{qnode}s by the users of its APIs.
The lock maintains both the \texttt{head} and \texttt{tail} pointers to the linked list formed by the \qnode{}s of waiting threads.
We refer the reader to~\cite{mcsk42} for full details.

In \uu{} of an MCS-K42 the following issues arise: 

\begin{description}
    \item[Mutex violation:] When there is already a thread in the \cs{} and an enqueued successor, a misbehaving thread, \misbehaver{} can release the successor into the \cs{} violating the mutual exclusion.
    \item[\misbehaver{} starvation:] When the lock is  unowned by any thread, the \misbehaver{} will starve until (potentially forever) the waiters' queue is formed .
    \item[Any thread starvation:]  Assume a thread  $T_1$ is in \cs{}. Let \misbehaver{} release the lock. When $T_1$ attempts its \rel{}, it would wait (potentially forever) until there is a successor .
\item [Stack corruption:]
  Consider two threads $T_1$ and $T_2$ simultaneously in the \cs{} caused by an instance of \uu{}.
  Eventually, both $T_1$ and $T_2$ may racily execute \rel{}. In a specific interleaving, $T_1$ may hold a pointer to $T_2$'s \qnode{}, while $T_2$ may have already finished its \rel{} and reclaimed its stack. In this situation, $T_1$ will write to and corrupt $T_2$'s stack.
 
\end{description}

\textbf{Detection and solution:}
We envision to re-purpose the fields of the lock \qnode{} structure to store the \pid{} of the lock owner to detect and remedy \uu{} in the MCS-K42 lock. 
The solution is involved: the \texttt{locked} field needs to be repurposed for the \pid{} when there are wainting threads and the \texttt{head} field needs to be repurposed when there are no waiters along with a dedicated bit to indicate whether the \texttt{head} is to be interpreted as \pid{}. We omit the details due to space constraints.

\subsection{Hemlock}
\begin{figure}
	\centering
	\includegraphics[scale=0.65]{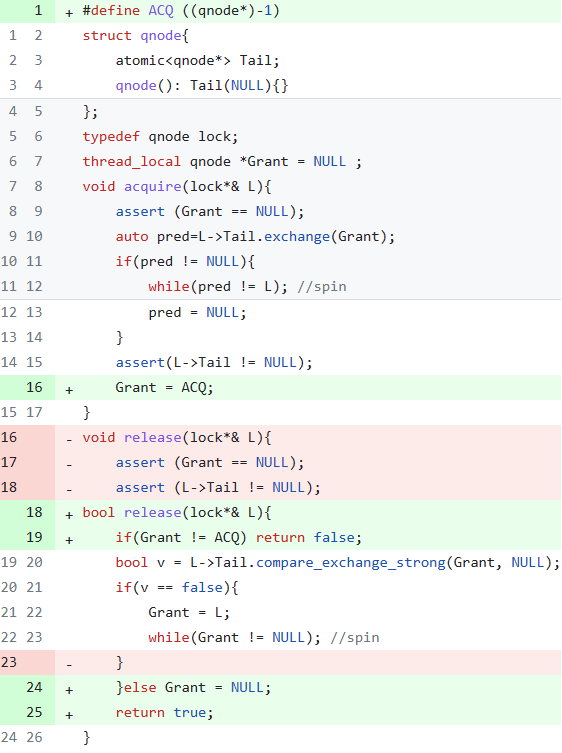}
	\caption{Hemlock before and after the fix.}
	\label{fig:Hemlock}
\end{figure}

Hemlock~\cite{hemlock} is the K42 counterpart for the CLH lock---it does away with carrying the context allocation and passing from its APIs. The lock protocol and remedy are shown in Figure~\ref{fig:Hemlock}. 

\textbf{Mutex violation and Starvation:} In an \uu{} situation, a misbehaving thread, \misbehaver{}, causes either assert violation at line 18 or starves self looping forever at line 22. As a result, the lock state is untouched and no mutex violation occurs. No starvation occurs for other threads. 

\textbf{Detection and solution:}  Our remedy to detect \uu{} in Hemlock is to set the \texttt{Grant} field with a special \texttt{ACQ} value in \texttt{acquire()}.  In the \texttt{release()}, we check if \texttt{Grant} is \texttt{ACQ} or \texttt{NULL}. A \texttt{NULL} value indicates \uu{}.
A successful \texttt{release()} protocol resets \texttt{Grant} to  \texttt{NULL}.

\subsection{Hierarchical locks} 
A series of multi-level locks were introduced to address remote memory accesses in Non-Uniform Memory Access (NUMA) systems. Here we discuss them informally.

\subsubsection{HMCS lock~\cite{chabbi2015high}}
\label{sec:hmcs}
The HMCS lock employs a tree of MCS locks mimicking the system's memory hierarchy.
A thread enters the \cs{} when it ``logically'' owns all locks from a leaf to the root of the tree.
Lock acquisition for each thread begins by competing for the local MCS lock at a designated leaf of the tree. 
If the thread is at the head of the MCS queue at a level, it proceeds to compete for the lock at the parent level.
A thread that acquires the lock at the root of the tree immediately enters the \cs{}.
Any thread that fails to immediately acquire a lock in the tree waits and receives that lock (and implicitly all ancestral locks) from its predecessor.
The release protocol also starts at a leaf of the tree.
A  waiting successor, if any, at the deepest node on a thread's path to the root implicitly inherits all locks held by its predecessor. As a result, when a successor is granted a lock, it immediately enters the \cs{} without explicitly contending for other locks along its path to the root.

In the event of an \uu{}, the HMCS lock suffers from the same set of issues as discussed previously in Section~\ref{sec:mcs} for the MCS lock.
Furthermore, since a lock releaser may proceed to release its ancestral MCS locks, it can introduce the same issue at every level in the tree.
Fortunately, the detection and remedy are also simple.
Since the release protocol starts at the leaf of the HMCS tree, we only need to adjust the MCS lock used at the leaf of the tree to use our renewed protocol discussed in the remedy for the MCS lock in Section~\ref{sec:mcs}. To be specific, the \acq{} protocol sets \texttt{I.locked} to \true{} after the lock acquisition, and the \rel{} protocol checks to make sure \texttt{I.locked} is \true{}; otherwise, it infers an instance of \uu{{} returning immediately.
At the end of a successful \rel{} protocol, \texttt{I.locked} is reset to \false{}.

The AHMCS lock~\cite{ChabbiAHMCS} is a refinement atop the HMCS lock allowing threads to start their \acq{} and the corresponding \rel{} at any level in the tree to dynamically adjust to contention. Our suggested remedy for HMCS applies to AHMCS as well since each thread brings its own \qnode{} allowing us to inspect whether the \texttt{locked} flag is set.

\subsubsection{HCLH lock~\cite{luchangco2006hierarchical}}
\label{sec:hclh}

The HCLH lock has two queues, a local queue (per NUMA domain) and a global queue. The thread at the head of the local queue waits to accumulate more requests within the local queue and then splices the local queue into the global queue via a \swap{} operation. 

A key deviation in HCLH over CLH is that predecessor \qnode{} over which a thread subsequently gains ownership, is returned as part of the \acq{} API instead of the \rel{} API (as was done in the CLH lock). 
As a result of this change, the \rel{} protocol in HCLH is minimal---just setting \texttt{I->succ\_must\_wait=false}.

In the event of an \uu{} in the HCLH lock, the  \qnode{} \text{I} used in \rel{} is not enqueued. Hence, changing \succwt{} on an un-enqueued node has no effect on the lock; consequently, there will be neither violation of the mutual exclusion nor starvation of any thread. Thus HCLH is relatively immune to \uu{}. We assume that the caller of \uu{} does not use any old \qnode{} on which previously it had ownership, which could put it back in the same situation as the CLH lock.

\subsubsection{HBO lock~\cite{RadovicHierarchicalBOLock}}
\label{sec:hbo}
The Hierarchical Backoff Lock (HBO) builds on the TAS lock discussed previously. 
In the HBO lock, the thread \cas{}es an ID representing its NUMA domain into the lock word \texttt{L} (instead of the \texttt{LOCKED} boolean).
Other threads can observe the NUMA domain ID of the lock holder when they attempt to acquire the lock.
With this knowledge, the threads from the same NUMA domain as the lock holder back off for a shorter period, whereas the threads from a remote NUMA domain back off for a longer period.

Our solution to detect and remedy \uu{} TAS lock can be retrofit for the HBO lock in the following way.
We need to \cas{} both the \pid{} of the lock owner as well as the NUMA domain ID of the lock owner into \texttt{L}, so that the \rel{} protocol can detect an \uu{} and the \acq{} protocol can detect how much to back off.
The two fields can be easily encoded (via bit shifting and bitwise-ORing) into a single 64-bit lock word---a 32-bit \pid{} and (say) an 8-bit NUMA domain ID.

\subsubsection{Cohort locks}
\label{sec:cohort}

Dice et al.~\cite{dice2012lock} invented cohort locks, which allows combining two locks (may be of different type) into a two-level hierarchy. The properties required to form a cohort lock out of two lock types $G$ and $S$ are: a) $G$ should allow the acquiring thread to differ from the releasing thread and b) $S$ should have the cohort detection property:
a thread releasing the lock can detect other threads are concurrently attempting to acquire the same lock. A cohort lock is formed by making $G$ a  global lock and by assigning each NUMA cluster $c$ with a distinct local lock $S_c$.

Dice et.al., discuss combining a local backoff lock (BO) with a global backoff lock to form a C-BO-BO lock. Combining two ticket locks (TKT) forms a C-TKT-TKT lock. Combining two MCS locks forms a C-MCS-MCS. Combining a global ticket lock with a local MCS lock forms C-TKT-MCS lock. The acquisition and release sequences always begin at the local lock and proceed to the global lock iff there is no cohort at the local level, similar to  HMCS.

In the event of an \uu{}, all these locks suffer from the issues that we already discussed for the corresponding locks used at the local level. As such, the detection and remedy are also obvious---reuse the techniques to detect and remedy the local lock. For example, by storing the \pid{} of the thread in the local BO lock (ref: Section~\ref{sec:tas}), we can fix a C-BO-BO lock; by storing the \pid{} of the local lock into a field of a ticket lock (ref: Section~\ref{sec:tas}), we can fix the C-TKT-TKT lock. C-MCS-MCS is a degenerate case of HMCS lock with 2 levels, and the solution trivially applies (ref: Section~\ref{sec:hmcs}). By employing the remedy discussed in Section~\ref{sec:mcs} for the MCS lock, C-TKT-MCS lock can be fixed.

\begin{table*}[]
\small
    \centering
    \begin{tabular}{r|c|c|c|l}
        Lock & violates  & starves  & starves  & detection + \\
        & Mutex & \misbehaver & others &   remedy \\ 
        \hline
        TAS & \cmark & \xmark & NA   & store \pid{} in \texttt{L} \\
        Ticket  & \cmark &  \xmark & \cmark &   introduce a new \pid{} field \\
        Anderson ABQL~\cite{anderson1990performance} & \cmark & \xmark &  \xmark  & check and reset \texttt{myPlace} in \rel{} \\
        Graunke-Thakker~\cite{GraunkeThakkarLock} & \xmark & \xmark & \cmark  & introduce \texttt{holder} array.\\
        MCS~\cite{tocs91} &  \cmark & \cmark & \xmark &  check \texttt{I.locked} and reset \texttt{I.next}\\
        CLH~\cite{magnusson1994queue, craig1993building} & \cmark & \xmark & \cmark  & check and reset \texttt{I.prev} in \rel{}\\
        MCS-K42~\cite{mcsk42} & \cmark & \cmark & \cmark & re-purpose \texttt{qnode}'s fields \texttt{locked} and \texttt{next} to store and manage lock owner's PID.\\ 
        Hemlock~\cite{hemlock} & \xmark & \cmark & \xmark & check and reset \texttt{Grant} in  \rel{}\\
        HMCS~\cite{chabbi2015high} & \cmark & \cmark & \xmark &  same as MCS at each level\\
        HCLH~\cite{luchangco2006hierarchical} & \xmark & \xmark & \xmark  & not applicable \\
        C-RW-NP/RP/WP~\cite{calciu2013numa} &\cmark & \xmark & \cmark  & unsolved\\
        Peterson’s lock~\cite{peterson1981myths} &\xmark & \xmark & \xmark  & not applicable \\
        Fisher’s lock\cite{lamport1987fast} &\cmark & \xmark & \xmark  & check and reset \texttt{x} in \rel{}\\
        Lamport’s Algo 1 / Algo 2~\cite{lamport1987fast}&\cmark & \xmark  & \cmark &  check and reset \texttt{y} in \rel{}\\
    \end{tabular}
    \caption{Analysis of popular locks considering a single misbehaving thread releasing the lock that it has not held.}
    \label{tab:locks_analysis}
\end{table*}

\subsection{Reentrant lock}
Java’s Open JDK reentrant lock~\cite{haack2008reasoning} implementation checks the owner before decrementing the count and hence is immune. However, the behavior may be different in other JDKs. Like in Open JDK, the Pthread implementation~\cite{pthread} of reentrant mutex records ownership in the \texttt{lock} operation and checks the owner in \texttt{unlock}. In an unbalanced-unlock scenario, \texttt{unlock} returns an error.

\section{Reader-Writer (RW) Locks}
\label{sec:rw}
A reader-writer lock allows the lock to be acquired in read-only mode in addition to the standard read-write mode. The read-only mode admits concurrent reader into \cs{}.
We assume four APIs in the context of RW locks: \rlock{} paired with \runlock{} used by readers and \wlock{} paired with \wunlock() used by the writers.
Invoking \runlock{} or~\wunlock{} without first invoking its corresponding locking API leads to an \uu{} situation. Mismatches such as pairing \rlock{}-\wunlock{} and \wlock{}-\runlock{} also cause \uu{}s.

Numerous RW-locking algorithms exist~\cite{MellorCurmmeyScalbelRW, dice2010tlrw, hsieh1991scalable, lev2009scalable, shirako2012design, calciu2013numa}, covering them all is  impractical. We focus on the recent NUMA-aware RW lock by Calciu et al~\cite{calciu2013numa}.
The NUMA-aware lock has several variants. We focus on the neutral-preference RW-lock (C-RW-NP), but the findings directly carry over to the other variants.

Figure~\ref{lst:rwnp} shows the C-RW-NP lock algorithm.
It uses a cohort lock~\cite{dice2012lock} and a scalable counter  (\texttt{ReadIndr}) as its building blocks.
 The cohort lock\cite{dice2012lock} employs a global partitioned ticket lock~\cite{dice2011brief} and a node-level ticket lock represented as C-PTK-TKT. 
 The  \texttt{ReadIndr} is a generic abstraction that allows \texttt{arrive} and \texttt{depart} APIs for readers to express their arrival and departure; an additional \texttt{isEmpty} API is provided to check for the absence of any readers by the writers. Underneath, \texttt{ReadIndr} may use SNZI~\cite{lev2009scalable},  per-NUMA domain counter, or split ingress-egress counters~\cite{calciu2013numa}.
 
 Both readers and writers compete for the same cohort lock as the first step of their \rlock{} and \wlock{} protocols, respectively.
 This ensures mutual exclusion between multiple writers and also between readers and a writer.
 After acquiring the cohort lock, the reader advertises its presence via \texttt{ReadIndr.arrive()} and immediately releases the cohort lock even before entering the \cs{}, which allows concurrent readers to enter the \cs{}. 
 The \runlock{} protocol simply  indicates the reader's departure via \texttt{ReadIndr.depart()}. 
 
 The writer after acquiring the cohort lock in \wlock{} waits until the concurrent readers, if any, in the \cs{} have drained, which is possible by querying \texttt{ReadIndr.isEmpty}. Notice that once a writer has acquired the cohort lock step in its \wlock{}, it obstructs subsequent readers (and writers) from acquiring the  cohort lock, thus ensuring its progress. 
 The \wlock{} is complete when \texttt{ReadIndr.isEmpty} is true,  admitting the writer into the \cs{}. The \wunlock{} API simply releases the cohort lock. Writers only query the \texttt{ReadIndr} and do not update it.

\begin{figure}[t]
\begin{minipage}{.45\linewidth}
\begin{lstlisting}[language=C++]
reader:
	CohortLock.acquire()
	ReadIndr.arrive()
	CohortLock.release()
 	<read-critical-section>
 	ReadIndr.depart()
\end{lstlisting}
\end{minipage}
\begin{minipage}{.45\linewidth}
\begin{lstlisting}[language=C++]
writer:
	CohortLock.acquire()
 	while (!ReadIndr.isEmpty())
 		Pause
 	<write-critical-section>
 	CohortLock.release()
\end{lstlisting}
\end{minipage}
\caption{C-RW-NP lock~\cite{calciu2013numa} algorithm.}
\label{lst:rwnp}
\end{figure}

\textbf{Mutex violation:} Consider a single reader (T$_r$) in the \cs{} and a writer (T$_w$)  waiting on \texttt{ReadIndr.isEmpty}; a misbehaving reader can make \texttt{ReadIndr} empty, letting both the T$_w$ and the T$_r$ simultaneously be in the \cs{} violating mutual exclusion.

\textbf{Starvation:} continuing with the previous example, subsequently, when T$_r$ exits \cs{}, it can corrupt the \texttt{ReadIndr}; a signed integer \texttt{ReadIndr} can become a negative value; an unsigned  \texttt{ReadIndr} can become a large positive value; a split counter can make ingress and egress counters diverge. 
In all these situations, subsequent writers may wait forever, spinning on \texttt{ReadIndr.isEmpty} in their \wlock{}. 
In addition, the C-RW-NP is susceptible to all the misuses resulting from \wunlock{} being invoked without first calling the \wlock{}, which has the same behavior as previously discussed in cohort locks~\ref{sec:cohort}.

\textbf{Detection and solution}: An unbalanced \wunlock{}, which uses the ticket lock, is detectable and remediable via the same strategy discussed in Section~\ref{sec:tkt}. 
An unbalanced \runlock{}, however, is not detectable or remediable without significant modifications to the read indicator.   
\texttt{ReadIndr} is compact and used for counting without maintaining reader's identity. 
We leave it for future research to address it in a scalable and memory-efficient manner. In Pthreads implementation of the RW lock interface, lock ownership information is maintained and the \unlock{} operation checks the caller thread's current mode (Read or Write or neither). As a result, \uu{} is detectable.


\section{Software-only locks} 
\label{sec:sw}
All the locks discussed previously depend on hardware-provided atomic operations. 
We now analyze software-only locks, which don't depend on special hardware support~\cite{peterson1981myths, lamport1987fast, yang1995fast}. We refer the readers to the Appendix added as auxiliary material for this paper. 

Peterson's lock~\cite{peterson1981myths} presents a solution to the {\em two-process} mutual exclusion.  The global variable \texttt{turn} is used as an arbiter when both the threads try to enter the \cs{}. Listing~\ref{lst:peterson} in Appendix shows the algorithm. In \uu{} scenario, let one of the two threads call \unlock{} to {\em reset} its flag (\texttt{flag[self]}) to indicate its unwillingness to enter \cs{}. As a result, it neither starves self or the other thread who may want to enter \cs{} nor does it violate mutual exclusion (as there are only two threads). Similar argument can be made when both threads misbehave, as both threads reset their respective flags.

Fisher's lock~\cite{lamport1987fast} is an N-process solution for the mutual exclusion problem. Listing~\ref{lst:Fisher} in Appendix shows the algorithm and the fix to make the lock \uu{} resilient. The operations in angle brackets are assumed to be atomic. The value of \texttt{x} is initialized to 0.  Suppose a thread T$_j$ is waiting to enter \cs{} while spinning on line 2 (also means that the current value of \texttt{x} is non-zero due to thread T$_i$ with id $i$ executing \cs{}). In \uu{} scenario, a lock release operation by thread T$_k$ sets \texttt{x} to zero, allowing T$_j$ to enter \cs{}. Thus allowing T$_i$ and T$_j$ in \cs{} simultaneously and violating mutual exclusion. A single instance of \uu{} lets at most one waiting thread into \cs{}. Also, a single instance of \uu{} does not starve the misbehaving thread or other threads. We propose a fix to make the lock resilient against \uu{}: the value of \texttt{x} is compared with the lock holder's ID, $i$, while exiting the critical section to detect and prevent the side effects of \uu{}.

In Lamport's N-process mutual exclusion algorithm~\cite{lamport1987fast}, Algorithm 1, a single instance of misuse could let another waiting thread into the \cs{} violating mutual exclusion: Let T$_i$ be in \cs{}. As a result, the current value of \texttt{x} and \texttt{y} are $i$. Suppose a misbehaving thread T$_m$ sets \texttt{y:=0} in \uu{} scenario. A third thread T$_j$ that wants to enter \cs{} will see that the conditions on lines 2, 4, and 5 are all false  and hence, will enter \cs{}. Thus, threads T$_i$ and T$_j$ will execute the \cs{} simultaneously. 

An \uu{} can also lead to starvation of another thread: suppose T$_i$ is about to execute line 5 and in the meanwhile T$_m$ executes line 8 (\texttt{y:=0}) in an \uu{} scenario. Now, T$_i$ sees that the condition on line 5 is not satisfied and hence, goes back to \texttt{start}. Lamport's Algorithm 2 exhibits similar behavior in \uu{} scenario as in Algorithm 1: the lock release protocol in Algorithm 2 sets \texttt{y:=0}, which is done in Algorithm 1 as well. In Algorithm 2, in addition, a process-specific flag \texttt{b[i]:=false} is written. This has no effect on other processes. As the usage of variable \texttt{y} in Algorithm 2 is similar to that in Algorithm 1, we observe that \uu{} scenario in Algorithm 2 has the same effects as that in Algorithm 1. Listings ~\ref{lst:LamAlg1} and ~\ref{lst:LamAlg2} in Appendix show these algorithms. In the proposed fix for both algorithms, the value of \texttt{y} is compared with the lock holder's ID, $i$, while exiting the critical section to detect and prevent the side effects of \uu{}.

\paragraph{Summary} Table~\ref{tab:locks_analysis} summarizes the analysis for all the locks just described. For each lock considered, the table shows if there is a violation of the mutual exclusion, starvation for the misbehaving thread, and starvation for other threads.
The table also shows, for each lock, if the misuse is detectable and if so, how it could be detected. If the misuse is preventable, the table also shows the necessary fix to the lock algorithm described previously.

By design some locks may require one thread to \acq{} and another thread to \rel{} the lock. To avoid flagging such a \rel{} as \uu{}, one can set an environment variable to disable the check for \uu{} in all our proposed remedies. Alternatively, a flag passed to the \rel{} API could serve the same purpose. We omit these details in the Figures shown previously for brevity.   

\section{Evaluation}
\label{sec:Evaluation}
We now present the evaluation of ten applications, each with six lock algorithms. We start with the evaluation methodology, which includes a brief discussion on the applications and the evaluation framework. This is followed by a comparative study for each application using the original and modified lock implementations. 

\paragraph{System configuration.}
We use a dual-socket system with a 24-core 2-way SMT Intel Xeon Gold 6240C@2.60GHz processor for a total of 48 hardware threads. 
The CPU has 64 KB shared data and instruction caches, 1 MB unified L2, and 36 MB L3 unified caches. The system has 384GB DDR4 memory running Rocky  Linux 9.

\paragraph{Benchmarks} We consider the following applications from SPLASH-2x~\cite{woo1995splash, parsec2011memo} and PARSEC 3.0~\cite{bienia11benchmarking}: $barnes$, $dedup$, $ferret$, $fluidanimite$, $fmm$, $ocean$, $radiosity$, $raytrace$, and $streamcluster$ with $Native$ input dataset.
These applications are considered because they are {\em lock-sensitive}. In addition, we create a synthetic application with empty critical section protected by calls to \texttt{omp\_set\_}- and \texttt{unset} lock.  
We evaluate each of the previously mentioned applications with the following locks: $TAS$, $Ticket$, $ABQL$, $MCS$, $CLH$, and $HMCS$.
 For the synthetic application, we measure the throughput in terms of number of calls (in million) to lock APIs per second. For other applications, we measure the execution time.
 
The modified locks are tested using the LiTL~\cite{litl} framework. LiTL is an open-source, POSIX-compliant framework that allows for the interception of calls to \texttt{pthread} mutex lock and condition variable APIs. All \texttt{pthread} mutex lock and condition variable related API calls in an application are mapped to corresponding APIs in LiTL implementations of 
 many lock algorithms from the literature. We modify the 
 lock implementations of previously mentioned lock algorithms in the LiTL framework to handle \uu{}.  

We start with a configuration where there is no contention for the lock (i.e., running only one thread) and go up to maximum contention for the lock (i.e., running threads utilizing all the physical cores on the machine). 
Every configuration of a test is run 5 times and best run of original lock implementation is compared with the best run of the modified lock implementation.


\paragraph{Results.} Table~\ref{tab:resiliente} shows the percentage overhead relative to LiTL implementations. We present execution times and throughput for all applications using various configurations of lock algorithms and thread combinations in the Appendix. In a majority of the experiments, the overhead of modified lock implementations is negligible. The overhead at the maximum number of threads (=48), is <5\% above the original implementation for all combinations of locks and applications considered except for TAS and Ticket locks for Radiosity, Streamcluster, Raytrace, and Synthetic applications. 
These applications are lock-intensive, and the fix involved to the lock protocols is a generic solution discussed in Section~\ref{sec:motivation}. 
Radiosity spends more than 25\% of the execution time at synchronization points~\cite{woo1995splash}. 
By design, TAS and ticket perform poorly under high lock contention and the added \pid{} field  exacerbates contention. 
Due to the fixes proposed for both TAS and Ticket locks, adding the \pid{} field, there is an additional \texttt{load} operation in the \texttt{release} API. This additional \texttt{load}  operation  
 causes the overhead  compared to the original \texttt{release} API, which involved only \texttt{store} operation. Fluidanimate and Streamcluster applications use \texttt{trylock}s, which is not available in the CLH lock; hence we omits those results.
 Fluidanimate and Ocean applications work only with the power-of-two threads and hence we show their results for 32 threads.
 Some numbers are negative, indicating that the performance differences are within a margin of measurement error.

\setlength{\tabcolsep}{0pt}

\begin{table}
   \small
        \begin{center}
            \begin{tabular}{r*{6}{|R|}}
            \diagbox[width=7.5em]{Application \\ (Threads)}{Lock} & \multicolumn{1}{|c}{TAS} & \multicolumn{1}{|c}{Ticket} &
            \multicolumn{1}{|c}{ABQL} & \multicolumn{1}{|c}{MCS} &
            \multicolumn{1}{|c}{CLH}  & \multicolumn{1}{|c|}{HMCS} \\[1em]
 
    Barnes (48)~\  & -0.1 & 1.04 & -0.1 & 0.54 & 0.93 & 1.18\\\hline
    Dedup (48)~\  & -1.6 & 3.47&-3.3&0.32 &4.5 & 1.62\\\hline
    Ferret (48)~\  &-0.3 & 0.42&-0.5&-0.1 & 1.45& -1.0\\\hline
    Fluidanimate (32)~\ & 0.19 & 2.80 & -0.8 & 1.96 & -0 &1.96\\\hline
    FMM (48)~\  & 0.00  & 0.64&-0.3&-0.9& 0.40 &-0.3\\\hline
    Ocean (32)~\  & 1.68 & 4.23 &3.79&0.94& 3.31 &0.55\\\hline
    Radiosity (48)~\ &2.08&19.5&0.87&2.62& 1.72 &-0.9\\\hline
    Raytrace (48)~\  & 16.9 & 86.7 & 3.08 & -0.9 & 2.83 &2.38\\\hline
    Streamcluster (48)~\  & 1.30 &61.3&1.72&1.13 & -0 &-2.2\\\hline
    Synthetic (48)~\ & 22.0 & 118. &-0.2&3.20 & 3.27&1.64\\\hline

        \end{tabular}
            \caption{Overhead (in \% ) due to fix applied to TAS, Ticket,  ABQL, MCS, CLH and HMCS locks for max. thread count.}
    \label{tab:resiliente}
    \end{center}
\end{table}

\section{Related Work}
\label{sec:related}
\paragraph{On lock design and performance}
There exists a rich and diverse body of work focusing on the design of efficient locks targeted at various hardware architectures. While a majority of these ~\cite{chabbi2015high, chabbi2017efficient, dice2012lock, tocs91, mcsk42, luchangco2006hierarchical, lozi2016fast, craig1993building, magnusson1994queue, GraunkeThakkarLock, anderson1990performance} rely on atomic instructions that all modern systems support, there are works that do not rely on such hardware support~\cite{peterson1981myths, lamport1987fast, yang1995fast, dijkstra1968cooperating}. In all these works, the design addresses the performance of lock considering factors such as contention, lock usage pattern---some threads reading mutable data and while others write---~\cite{lev2009scalable, calciu2013numa, MellorCurmmeyScalbelRW}, fairness criteria~\cite{anderson1990performance, GraunkeThakkarLock, reed1979synchronization, tocs91}, energy efficiency~\cite{falsafi2016unlocking}, whether running on single core or multicore~\cite{glibcnptl}, and NUMA hierarchy~\cite{chabbi2015high, chabbi2017efficient, dice2012lock, calciu2013numa, luchangco2006hierarchical}. In comparison, we consider {\em unintentional} misuse and the resilience of locks against misuse rather than the performance of locks.

Guerraoui et al.~\cite{guerraoui2019lock} do a thorough empirical evaluation of existing lock algorithms with a broad range of applications on a variety of architectures. Also, they provide a simple and effective way to evaluate new lock algorithms~\cite{litl}. While they suggest guidelines for software developers to overcome performance bottlenecks when using locks, they do not consider the misuse scenarios. We use their framework~\cite{litl} to evaluate our modified lock implementations. 
\paragraph{On concurrency bugs and support system} The {\uu}  scenario is a specific case of a concurrency bug. Race, deadlock detection, atomicity violation, and techniques to uncover concurrency bugs dynamically and statically is the focus in some of the important works~\cite{lu2008learning, choi2002efficient, engler2003racerx, prvulovic2003reenact, savage1997eraser, yu2005racetrack, boyapati2002ownership, lucia2011isolating, ZihengGoAsplos, jin2012automated, jula2008deadlock, li2019efficient, sen2008race}. 
Part of the work presented in this paper focuses on the prevention of a specific concurrency bug, {\uu}, through better engineering of lock designs. From a software engineering perspective, by making the unbalanced-unlock usage side effect free, our proposed fixes improve the software reliability in mission-critical systems and do not compromise the entire system due to one misbehaving thread.
More elaborate feedback to the programmer via returning error codes or raising panics is also a possible design choice:  e.g., \texttt{pthread\_mutex} lock implements a separate lock type \texttt{PTHREAD\_MUTEX\_ERRORCHECK}, which allows an error code to be returned when a thread calls \unlock without holding the lock. Golang introduces API support to panic when an unlocked \texttt{sync.Mutex} is unlocked. The C++ thread sanitizer built into the  LLVM toolchain also takes the same approach as that of Golang. However, the panic is enabled only in debug mode. In comparison to these approaches, 
 our proposed fixes modify the lock protocol and hence, do not affect the application code.
\section{Conclusions}
\label{sec:conclusion}
In this paper, we presented an approach to making popular locking algorithms resilient to accidental misuse. 
We considered a specific misuse scenario, \uu{}, where a misbehaving thread calls \unlock without holding the lock. 
We presented data to show that \uu{} is a surprisingly common problem in large popular open-source repositories.
A systematic analysis of popular locks in the presence of  \uu{} showed that the misuse could violate mutual exclusion, lead to starvation, corrupt the internal state of the lock and/or thread, and sometimes be side-effect free. We then presented remedies to avoid the side effects, and applied these remedies to a representative set of lock implementations. We evaluated the modified lock implementations for a set of lock-sensitive applications and showed that the modified lock implementations do not significantly affect the performance.   

\clearpage
\bibliographystyle{ACM-Reference-Format}
\balance
\bibliography{paper}
\clearpage
\appendix
\begin{center}{\LARGE Appendix } \end{center}

\bigskip
We present the listings of four software-only lock algorithms, which are analyzed for \uu{} resilience in the paper.  
For all other locks analyzed in the paper, we present here detailed experimental results showing the percentage overhead of modified lock implementations w.r.t. implementations of original lock algorithms in LiTL framework.
The results show the overhead for different combinations of thread count, lock protocol, and application.

\section{Software-only  Locks}

\begin{figure}[h]
\begin{lstlisting}[language=C++, caption=Peterson's algorithm for 2 threads~\cite{arpaci2018operating}.,label=lst:peterson]
int flag[2];
int turn;
void init() {
    // indicate you intend to hold the lock with 'flag'
    flag[0] = flag[1] = 0;
    // whose turn is it? (thread 0 or 1)
    turn = 0;
}

void acquire() {
    // 'self' is the thread ID of caller
    flag[self] = 1;
    // make it other thread's turn
    turn = 1 - self;
    while ((flag[1-self] == 1) && (turn == 1 - self)); // spin-wait while it's not your turn
}

void release() {
    // simply undo your intent
    flag[self] = 0;
}

\end{lstlisting}
\end{figure}


\begin{figure}[h]
\begin{minipage}{.46\linewidth}
\begin{lstlisting}[language=C++,escapechar=&]
start:
    while <x != 0>;
        <x := i>;
        <delay> 
    if <x != i> goto start;
    critical section;
    &\colorbox{pink}{        }&
    x := 0
exit:
\end{lstlisting}
\end{minipage}
\begin{minipage}{.46\linewidth}
\begin{lstlisting}[language=C++,escapechar=&]
start:
    while <x != 0>;
        <x := i>;
        <delay> 
    if <x != i> goto start;
    critical section;
    &\colorbox{green}{if <x != i> goto exit;}&
    x := 0
exit:
\end{lstlisting}
\end{minipage}
\caption{Fischer’s algorithm before and after the fix~\cite{lamport1987fast}.}
\label{lst:Fisher}
\end{figure}


\begin{figure}[h]
\begin{minipage}{.45\linewidth}
\begin{lstlisting}[language=C++,escapechar=& ]
start:  <x := i> ;
    if  <y != 0> then goto start fi;
    <y := i> ;
    if  <x != i>  then delay;
        if < y != i > then goto start fi;
    fi
    critical section;
    &\colorbox{pink}{        }&
    <y := 0> 
exit:
\end{lstlisting}
\end{minipage}
\begin{minipage}{.45\linewidth}
\begin{lstlisting}[language=C++,escapechar=&]
start:  <x := i> ;
    if  <y != 0> then goto start fi;
    <y := i> ;
    if  <x != i>  then delay;
        if < y != i > then goto start fi;
    fi
    critical section;
    &\colorbox{green}{if <y != i> goto exit;}&
    <y := 0>
exit:
\end{lstlisting}
\end{minipage}
\caption{Lamport's algorithm 1 before and after the fix~\cite{lamport1987fast}.}
\label{lst:LamAlg1}
\end{figure}

\begin{figure}[h]
\begin{minipage}{.45\linewidth}
\begin{lstlisting}[language=C++,escapechar=&]
start:<b[i] := true> ;
    <x := i> ;
    if  <y != 0> then  <b[i] := false>;
        await  <y = 0>;
        goto start 
    fi;
    <y := i> ;
    if  <x = i>  then  <b[i] := false>;
        for j := 1 to N do await <NOT b[j]> od;
        if  <y != i>  then await  <y = 0>;
		goto start 
        fi 
    fi;
    critical section;
    &\colorbox{pink}{                  }&
    <y := 0> ;
    <b[i] := false> 
exit:
\end{lstlisting}
\end{minipage}
\begin{minipage}{.45\linewidth}
\begin{lstlisting}[language=C++,escapechar=&]
start:<b[i] := true> ;
    <x := i> ;
    if  <y != 0> then  <b[i] := false>;
        await  <y = 0>;
        goto start 
    fi;
    <y := i> ;
    if  <x = i>  then  <b[i] := false>;
        for j := 1 to N do await <NOT b[j]> od;
        if  <y != i>  then await  <y = 0>;
		goto start 
        fi 
    fi;
    critical section;
    &\colorbox{green}{if <y = i> then goto exit fi}&
    <y := 0> ;
    <b[i] := false> 
exit:
\end{lstlisting}
\end{minipage}
\caption{Lamport's algorithm 2 before and after the fix~\cite{lamport1987fast}.}
\label{lst:LamAlg2}
\end{figure}

\subsection{Lamport's Bakery algorithm}
 In the event of an unbalanced-unlock, Lamport's Bakery algorithm~\cite{bakery} shows no mutex violation and does not starve any thread including \misbehaver{}. The \rel{} protocol resets the lock holder's \texttt{number[i]} to zero. This reset has no side effect on any thread. 

\section{Detailed Overhead Analysis.}
Figure~\ref{fig:overhead} shows the percentage overhead of various locks on different applications at different concurrency level. Also included is the overhead over the range of threads per application. In each setting, we compare the lock with our proposed modifications for handling \uu{} against the original implementations. The `\#' mark shown in Figure~\ref{fig:overhead} represents no runs for 48 threads for all the locks in Fluidanimate and Ocean due to application requirement that power-of-two thread count is used in the runs. Fluidanimate and Streamcluster applications use \texttt{trylock}s, which is not available in the CLH lock. Hence, corresponding entries are marked `*' in Figure~\ref{fig:overhead}.

\begin{figure*}
    \centering
    \includegraphics[width=.8\linewidth]{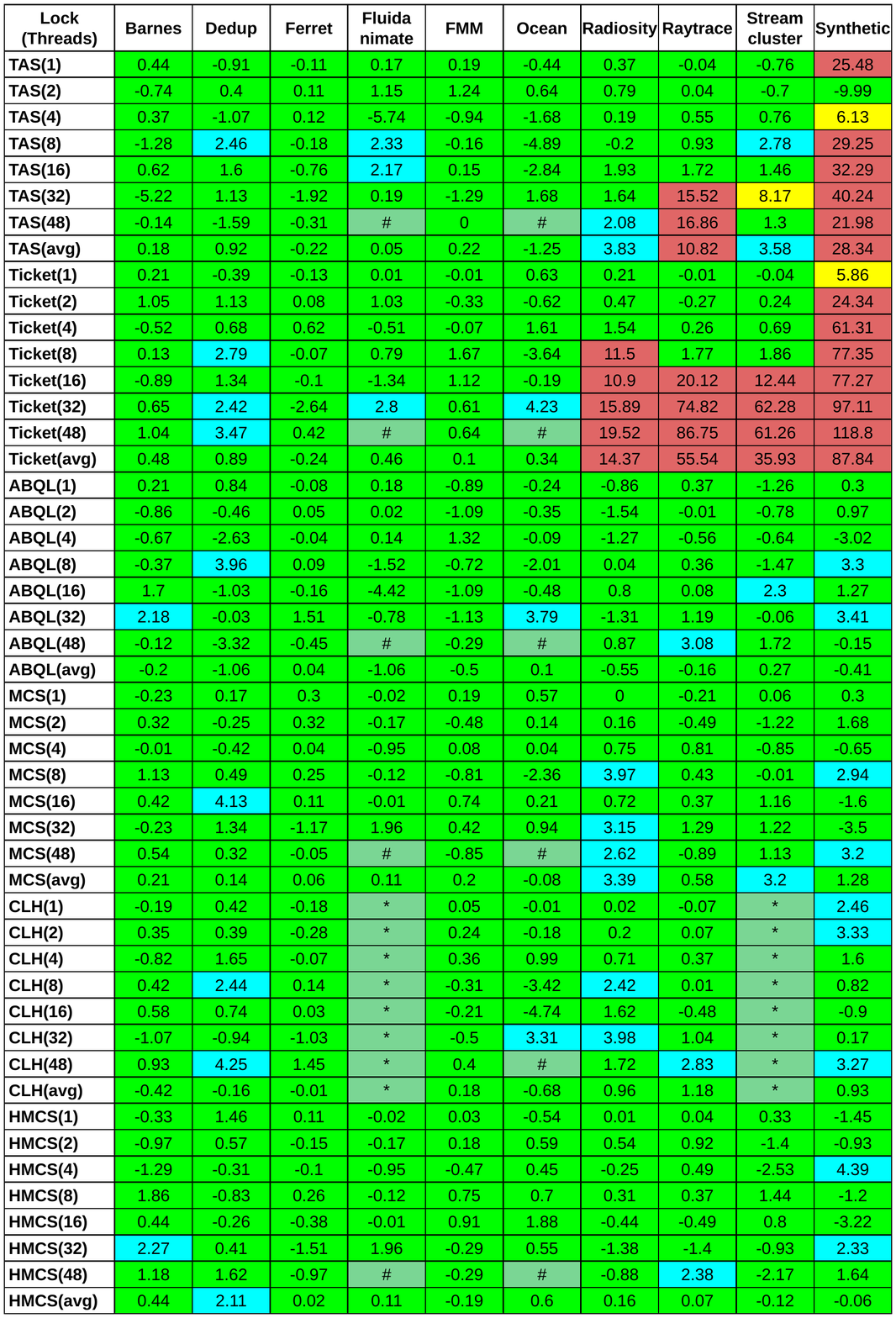}
			\caption{Percentage (\%) overhead of various locks on different applications at different number of threads. Baseline is the  original lock implementation. }
	\label{fig:overhead}
\end{figure*}

\clearpage

\end{document}